\documentclass[pra,aps,10pt,twocolumn,floatfix]{revtex4-1}
\usepackage{graphicx,color}
\usepackage{amsmath,amssymb,bm}
\usepackage{mathtools}
\usepackage{simplewick}
\usepackage{braket}		

\usepackage[plainpages=false,pdfpagelabels,colorlinks=true,linkcolor=red,urlcolor=blue,citecolor=blue,pdftitle={Entanglement Entropy of Eigenstates of Quantum Chaotic Hamiltonians},pdfauthor={},pdfdisplaydoctitle=true,pdfduplex=DuplexFlipLongEdge]{hyperref}

\definecolor{darkred}{rgb}{0.90,0.2,0.2}
\definecolor{darkgreen}{rgb}{0,0.60,.2}
\definecolor{darkblue}{rgb}{0.1,0.3,1}
\definecolor{grey}{cmyk}{0,0,0,0.25}
\definecolor{orange}{cmyk}{0,0.6,0.8,0}

\begin{document}
\title{Entanglement Entropy of Eigenstates of Quantum Chaotic Hamiltonians}

\author{Lev Vidmar}
\author{Marcos Rigol}
\affiliation{Department of Physics, The Pennsylvania State University, University Park, PA 16802, USA}

\begin{abstract}
In quantum statistical mechanics, it is of fundamental interest to understand how close the bipartite entanglement entropy of eigenstates of quantum chaotic Hamiltonians is to maximal. For random pure states in the Hilbert space, the average entanglement entropy is known to be nearly maximal, with a deviation that is, at most, a constant. Here we prove that, in a system that is away from half filling and divided in two equal halves, an upper bound for the average entanglement entropy of random pure states with a fixed particle number and normally distributed real coefficients exhibits a deviation from the maximal value that grows with the square root of the volume of the system. Exact numerical results for highly excited eigenstates of a particle number conserving quantum chaotic model indicate that the bound is saturated with increasing system size.
\end{abstract}

\maketitle

\emph{Introduction.--}
Entanglement in few-body quantum systems is a topic that triggered special interest since the birth of quantum mechanics~\cite{epr_35, schroedinger_35}. In recent years, interest has shifted towards entanglement in systems with many degrees of freedom, which, e.g., is relevant to current problems in the fields of condensed matter, quantum information, and quantum gravity~\cite{amico_fazio_08}. In condensed matter and quantum information, the concept of entanglement has played an essential role in designing efficient numerical algorithms~\cite{white_92, schollwoeck_05, schollwoeck_11}, in understanding quantum phase transitions~\cite{osterloh_amico_2002, osborne_nielsen_02, gu_deng_04}, and in characterizing the dynamics after quantum quenches~\cite{calabrese_cardy_05, eisler07, calabrese_cardy_07}. Also, studies with ultracold atoms in optical lattices~\cite{islam_ma_15, kaufman_tai_16} have begun the experimental exploration of entanglement in and out of equilibrium. An important aspect of these experimental systems is that the size of the subsystem of interest is not necessarily a vanishing fraction of the size of the entire system, in contrast to traditional statistical mechanics.

Despite considerable theoretical efforts~\cite{znidaric07a, deutsch_10, santos_polkovnikov_12, hamma_santra_12, vinayak_znidaric_12, deutsch13, alba15, beugeling15, yang_chamon_15, vivo_pato_16, dora_lovas_17, garrison15, dymarsky_laskhari_17, fujita_nakagawa_17}, rigorous understanding of the behavior of the bipartite entanglement entropy in eigenstates of generic (quantum chaotic) Hamiltonians (with ground states being an exception~\cite{eisert_cramer_10, srednicki_93, vidal_latorre_03, calabrese_cardy_04, hastings_07}) is lacking. The expectation is that typical eigenstates at high temperature are (nearly) maximally entangled. This follows from the result by Page~\cite{page93}, who proved that, for a bipartition of a system into subsystem $A$ and its complement $B$, the average entanglement entropy of random pure states is
\begin{equation} \label{def_Page}
 S_{\rm ave} = \ln {\cal D}_A  - \frac{1}{2} \frac{{\cal D}_A^2}{\cal D} \, ,
\end{equation}
where ${\cal D}$ and ${\cal D}_A$ (${\cal D}_{\rm A}\leq\sqrt{\cal D}$) are the Hilbert-space dimensions of the system and the subsystem $A$, respectively. Page's result suggests that if the ratio $f$ (referred to as the {\it subsystem fraction}) between the volume of subsystem $A$ and of the system is $f<1/2$, then the deviation from the maximum entanglement entropy vanishes exponentially with the volume of the system, while for $f=1/2$ the deviation is 1/2. However, eigenstates of physical Hamiltonians are measure zero in the space of pure states so one might argue that the previous expectation is ill founded. Indeed, eigenstates of translationally invariant quadratic fermionic Hamiltonians have been proved to violate Eq.~\eqref{def_Page}~\cite{vidmar_hackl_17}.

Here, we study the bipartite von Neumann entanglement entropy (referred to as the {\it entanglement entropy}) of pure states with a fixed particle number (towards the end, we briefly explore what happens when this constraint is lifted). We consider random pure states with normally distributed real coefficients (referred to as {\it random canonical states}), which are motivated by the Gaussian orthogonal ensemble of random matrix theory \cite{dalessio_kafri_16}, and high-energy eigenstates of a particle number conserving quantum chaotic model of hard-core bosons. We show that the average entanglement entropy of random canonical states consists of two terms: (i) a ``mean-field'' term associated with the maximum entanglement entropy, and (ii) a fluctuation term arising from fluctuations of the matrix elements of the reduced density matrix. In general, the latter cannot be neglected in finite systems. In particular, in a system that is away from half filling and divided in two equal halves, we prove that an upper bound to the average entanglement entropy of random canonical states exhibits a deviation from the maximal value that grows with the square root of the subsystem volume. Numerical results for those states, and for high-energy eigenstates of the hard-core boson model, indicate that the bound is saturated with increasing system size.

{\it Quantum chaotic model.--}
We consider hard-core bosons in one-dimensional lattices with nearest and next-nearest neighbor hoppings ($t_1$ and $t_2$) and interactions ($V_1$ and $V_2$)
\begin{eqnarray} \label{def_Ham}
 \hat H & = & - t_1 \sum_{l=1}^{L} (\hat b_{l+1}^\dagger \hat b^{}_l + {\rm H.c.}) - t_2 \sum_{l=1}^{L} (\hat b_{l+2}^\dagger \hat b^{}_l + {\rm H.c.}) \nonumber \\ &  & + V_1 \sum_{l=1}^L  \hat n_l \hat n_{l+1} + V_2 \sum_{l=1}^L \hat n_l \hat n_{l+2} \, ,
\end{eqnarray}
where $\hat n_l = \hat b_l^\dagger \hat b_l$, $(\hat b_l)^2 = (\hat b_l^\dagger)^2 = 0$, and $L$ is the number of lattice sites. In our calculations, we set $t_1=t_2 = 1$ and $V_1=V_2=1.1$. For these parameters, this model has been shown~\cite{rigol_09a, santos_rigol_10a} to be quantum chaotic and exhibit eigenstate thermalization~\cite{deutsch_91, srednicki_94, rigol_dunjko_08, dalessio_kafri_16} for the system sizes studied here. We use exact diagonalization, resolving all symmetries, to compute the average entanglement entropy $\bar S$ of eigenstates in the center of the spectrum (to reduce finite-size effects, we only consider 20\% of all eigenstates). For an eigenstate $|\Psi\rangle$, the entanglement entropy is $S = -{\rm Tr}\{ \hat\rho_A \ln(\hat\rho_A) \}$, where $\hat \rho_A = {\rm Tr}_B\{ |\Psi \rangle \langle \Psi | \}$ is obtained from the spatial trace over the degrees of freedom in subsystem $B$.

{\it Random canonical states.--}
We construct random canonical states on a lattice with $L$ sites and two states per site as $|\psi_N\rangle = \sum_{j=1}^{{\cal D}_N} z_j | j \rangle/\sqrt{{\cal D}_N}$, where $z_j$ is a normally distributed real random number with zero mean and variance one, $N$ is the particle number, ${\cal D}_N = \binom{L}{N}$ is the dimension of the Hilbert space, and $|j\rangle$ is a base ket for $N$ particles in the site-occupation basis. Note that, in finite systems, $|\psi_N\rangle$ is not exactly normalized. However, for the normalized state $|\psi_N\rangle/\sqrt{\cal N}$, the mean of the normalization factor ${\cal N}$ is one, and its fluctuations vanish exponentially fast with increasing $L$ (see Ref.~\cite{suppmat}). This justifies the use of $|\psi_N\rangle $ in the analytical calculations.

We consider a bipartition into subsystem $A$ (with $L_A<L$ consecutive lattice sites) and its complement subsystem $B$ (with $L-L_A$ lattice sites). One can express $|\psi_N\rangle$ as a sum of direct products of base kets in subsystems $A$ and $B$,
\begin{equation} \label{def_psiN}
 |\psi_N\rangle = \sum_{N_A=N_A^{\rm min}}^{N_A^{\rm max}} \sum_{a=1}^{d_{N_A}} \sum_{b=1}^{d_{B(N_A)}} \frac{z_{a,b}(N_A)}{\sqrt{{\cal D}_N}}|a,N_A\rangle |b,N-N_A \rangle \, .
\end{equation}
In the latter expression, the states in subsystem $A$ can be seen to belong to sectors with different particle number $N_A$, where $N_A^{\rm min} = {\rm Max}[0,N-(L-L_A)]$ and $N_A^{\rm max} = {\rm Min}[N,L_A]$. The Hilbert-space dimension of a sector with $N_A$ particles is $d_{N_A} = \binom{L_A}{N_A}$. For every sector in subsystem $A$ with $N_A$ particles, the corresponding sector in subsystem $B$ contains $N-N_A$ particles and has a Hilbert-space dimension $d_{B(N_A)} = \binom{L-L_A}{N-N_A}$.

The reduced density matrix of subsystem $A$, $\hat \rho_A = {\rm Tr}_B\{ |\psi_N \rangle \langle \psi_N | \}$, can be written as
\begin{equation} \label{def_rhoA}
 \hat \rho_A = \sum_{N_A=N_A^{\rm min}}^{N_A^{\rm max}} \sum_{a, a'= 1}^{d_{N_A}} |a,N_A \rangle \langle a',N_A | \frac{F(a,a',N_A)}{{\cal D}_N} \, ,
\end{equation}
where $\hat \rho_A$ is block diagonal with each block labeled by $N_A$, and 
\begin{equation}
 F(a,a',N_A) = \sum_{b=1}^{d_{B(N_A)}} z_{a,b}(N_A) z_{a',b}(N_A) \, ,
\end{equation}
which is a sum of products of random numbers, and whose average is $\overline{F(a,a',N_A)} = d_{B(N_A)} \delta_{a,a'}$. Hence, the average reduced density matrix is
\begin{equation} \label{def_rhoA_bar}
 \hat{\bar\rho}_A = \sum_{N_A=N_A^{\rm min}}^{N_A^{\rm max}} \sum_{a = 1}^{d_{N_A}} |a,N_A \rangle \langle a,N_A | \, \bar \lambda_{N_A} \, ,
\end{equation}
i.e., it is diagonal with diagonal matrix elements $\bar \lambda_{N_A} = d_{B(N_A)}/{{\cal D}_N}$.

\begin{figure}[!t]
\begin{center}
\includegraphics[width=0.98\columnwidth]{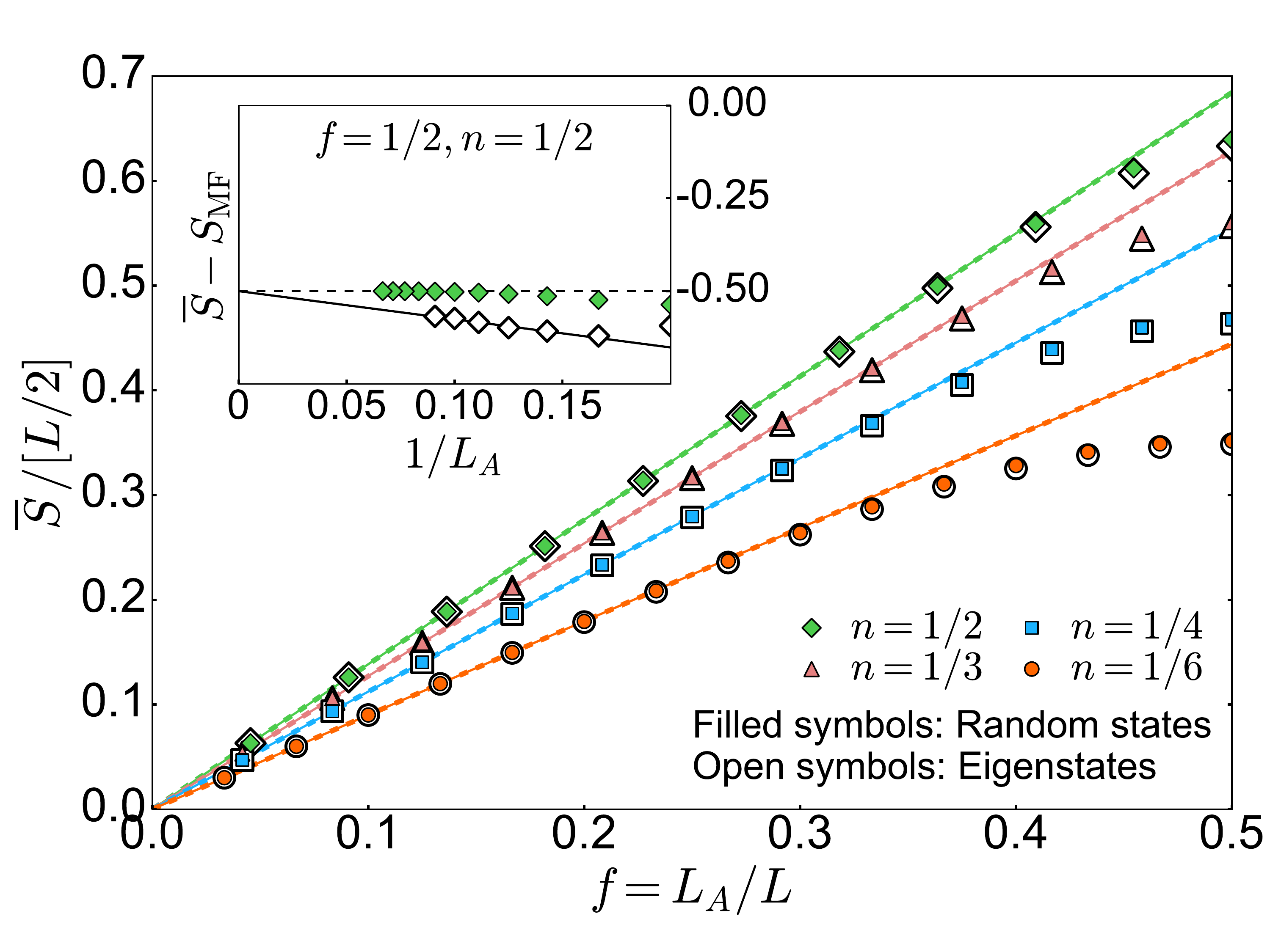}
\caption{Average entanglement entropy in eigenstates of a quantum chaotic Hamiltonian, Eq.~(\ref{def_Ham}), and in random canonical states, Eq.~(\ref{def_psiN}). (Main panel) $\bar S/[L/2]$ vs the subsystem fraction $f=L_A/L$ for $n=1/2$, 1/3, 1/4 and 1/6 ($L=22$, 24, 24, and 30, respectively). Solid (dashed) lines show the mean-field entanglement entropy $S_{\rm MF}$ ($S_{\rm MF}^*$) from Eq.~(\ref{Smf_long}) [Eq.~(\ref{Smf_short})]. (Inset) $\bar S - S_{\rm MF}$ vs $1/L_A$ for $f=n=1/2$.} 
\label{fig1}
\end{center}
\end{figure}

{\it Entanglement entropy of random canonical states.--}
We are interested in the average of the entanglement entropy over random canonical states, $\bar S = - \overline{{\rm Tr}\{ \hat\rho_A \ln(\hat\rho_A) \}}$. To compute it, we define the operator 
\begin{equation}
 \hat M = (\hat{\bar\rho}_A)^{-1} (\hat\rho_A - \hat{\bar\rho}_A) \, ,
\end{equation}
so that $S$ of a single random state can be written as
\begin{equation} \label{def_S_vs_M}
 S = - {\rm Tr}\left\{ \hat{\bar\rho}_A (\hat I + \hat M) \ln \left[ \hat{\bar\rho}_A (\hat I + \hat M) \right] \right\} \, .
\end{equation}
$\hat{\bar\rho}_A$ and  $\hat I + \hat M$ commute for random canonical states as, in the site-occupation basis, $\hat I + \hat M$ is block diagonal with matrix elements $[I + M(N_A)]_{a,a'} = F(a,a',N_A)/d_{B(N_A)}$. Hence, the logarithm in Eq.~\eqref{def_S_vs_M} can be replaced by a sum of logarithms, and the entanglement entropy can be written as $S = S_{\rm MF} + S_0 + S_{\rm fluct}$, where
\begin{eqnarray}
 S_{\rm MF} & = & - {\rm Tr}\left\{ \hat{\bar\rho}_A \ln \hat{\bar\rho}_A \right\} \label{def_Smf}, \\ S_0 & = & - {\rm Tr}\left\{ \hat{\bar\rho}_A  \hat M \ln \hat{\bar\rho}_A \right\} \label{def_S0}, \\ S_{\rm fluct} & = & - {\rm Tr}\left\{ \hat{\bar\rho}_A (\hat I + \hat M) \ln (\hat I + \hat M) \right\} \, . \label{def_Sfluct}
\end{eqnarray}
We call $S_{\rm MF}$ in Eq.~(\ref{def_Smf}) the ``mean-field'' entanglement entropy as the elements of the reduced density matrix in Eq.~(\ref{def_rhoA}) are replaced by their average [see Eq.~(\ref{def_rhoA_bar})]. As a result, $\bar S_{\rm MF} = S_{\rm MF}$.
The terms in Eqs.~(\ref{def_S0})-(\ref{def_Sfluct}) contain the contribution to the entanglement entropy due to fluctuations of the matrix elements of the reduced density matrix about their average. Since $\overline{M(N_A)_{a,a'}} = 0$, the average of $S_0$ is zero, $\bar S_0 = - \sum_{N_A} \bar\lambda_{N_A} \ln \bar\lambda_{N_A} \sum_{a} \overline{M(N_A)_{a,a}} = 0$. Hence, only $S_{\rm fluct}$ is nontrivial. Since $\bar S_{\rm fluct} \neq 0$ in general, it then follows that $\overline{{\rm Tr}\{ \hat\rho_A \ln(\hat\rho_A) \}} \neq {\rm Tr}\{ \hat{\bar\rho}_A \ln(\hat{\bar\rho}_A)\}$.

\begin{figure*}[!t]
\begin{center}
\includegraphics[width=1.99\columnwidth]{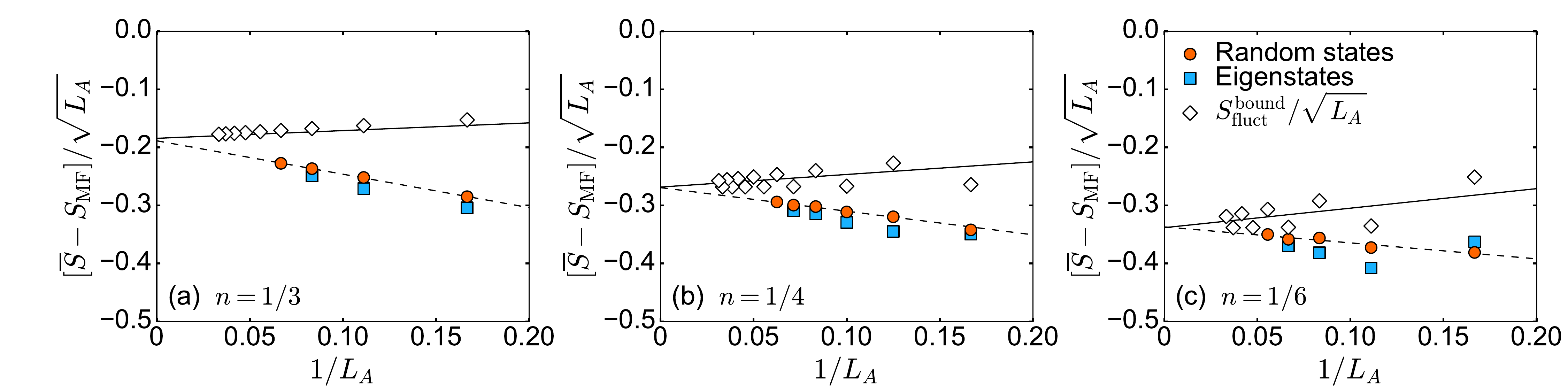}
\caption{Fluctuation contribution to the average entanglement entropy when $f=1/2$, for average site occupations $n=1/3$ (a), 1/4 (b), and $1/6$ (c). The circles display results for random canonical states, Eq.~(\ref{def_psiN}) (see Ref.~\cite{suppmat} for details on the numerical calculations), and the dashed lines are linear fits to those results. The squares display the results for eigenstates of Hamiltonian~(\ref{def_Ham}). The diamonds are the upper bound $S^{\rm bound}_{\rm fluct}$ from Eq.~(\ref{def_Sbound}) and the solid lines are $S^{\rm bound*}_{\rm fluct}$ from Eq.~(\ref{def_Sstar}). The zigzag structure of $S^{\rm bound}_{\rm fluct}$ in (b) and (c) is a finite-size effect~\cite{suppmat}.}
\label{fig2}
\end{center}
\end{figure*}

{\it Mean-field entanglement entropy.--}
We first review the key properties of $S_{\rm MF}$ in Eq.~(\ref{def_Smf})
\begin{equation}
S_{\rm MF} = - \sum_{N_A=N_A^{\rm min}}^{N_A^{\rm max}} d_{N_A} \bar\lambda_{N_A} \ln \bar\lambda_{N_A} \, . \label{Smf_long}
\end{equation}
Using Stirling's approximation for large systems, the two leading-order terms of Eq.~(\ref{Smf_long}) can be written as~\cite{suppmat}
\begin{equation} \label{Smf_short}
 S_{\rm MF}^* =  -L_A  \left[  n \ln n + (1-n) \ln(1-n) \right]+ \frac{f + \ln(1-f)}{2}\, ,
\end{equation}
where $n=N/L$ is the average site occupation and $f=L_A/L$ is the subsystem fraction. The leading term in Eq.~(\ref{Smf_short}) is identical to the one derived in Ref.~\cite{garrison15}. Two notable features of $S_{\rm MF}^*$ are: (i) it is proportional, up to nonextensive corrections, to the subsystem volume $L_A$, and (ii) the proportionality constant only depends on $n$.

Figure~\ref{fig1} (main panel) compares the average entanglement entropy of the eigenstates in the center of the spectrum (20\% of all eigenstates) of Hamiltonian~(\ref{def_Ham}), and the average entanglement entropy of random canonical states introduced in Eq.~(\ref{def_psiN}). The results are plotted vs the subsystem fraction $f$ for different values of $n$, and yield a remarkable agreement. The solid and dashed (overlapping) lines show the mean-field entanglement entropies $S_{\rm MF}$ and $S_{\rm MF}^*$ from Eqs.~(\ref{Smf_long}) and~(\ref{Smf_short}), respectively. One can see that $S_{\rm MF}$ is close to $\bar{S}$ for small $f$, but the two depart as $f \to 1/2$. The inset in Fig.~\ref{fig1} indicates that, as $L_A\rightarrow\infty$ for $f=1/2$ and $n=1/2$, the deviation from $S_{\rm MF}$ is the one predicted by Page [Eq.~(\ref{def_Page})] upon replacing $\ln {\cal D}_A$ with $S_{\rm MF}$ (because of the fixed particle number), i.e., $\bar S = S_{\rm MF} - 1/2$ for random canonical states and for the Hamiltonian eigenstates. Surprisingly, for all other average site occupations ($n\neq1/2$) at $f=1/2$, we find that the deviation from $S_{\rm MF}$ is not $O(1)$. The main goal of this Letter is to understand this deviation.

{\it Fluctuation term.--}
Next, we focus on $S_{\rm fluct}$ in Eq.~(\ref{def_Sfluct}). It is key to understand the distribution of the eigenvalues $\Lambda_j$ of the block-diagonal matrix $I + M$. Even though the average of this matrix is the identity matrix, some of its eigenvalues can strongly deviate from one. As we show next, this occurs when the total number of particles (or holes, if $N > L/2$) is smaller than the subsystem volume, and follows from the fact that the eigenvalues within each sector with particle number $N_A$ satisfy the sum rule $\overline{{\rm Tr} \{ I+ M(N_A) \}} = d_{N_A}$. 

Without loss of generality, we focus on site occupations $n < 1/2$ [if $n>1/2$, one should focus on the hole occupations $n_h = 1-n$, since $\bar S(n) = \bar S(1-n)$]. Let us first consider the sector with $N_A=N$. In that case, one can write $I + M(N) = \vec{z}_1(N) \vec{z}_1(N)^{T}$, where $\vec{z}_b(N_A) \coloneqq (z_{1,b},z_{2,b},...,z_{d_{N_A},b})^T$ is a random vector. Hence, $I + M(N)$ has rank one and therefore contains only one nonzero eigenvalue. In average, this nonzero eigenvalue equals $d_N = \binom{L_A}{N}$.

Moving on to sectors with an arbitrary number of particles $N_A$, one can write
\begin{equation}  \label{def_I_M}
 I + M(N_A) = \frac{1}{d_{B(N_A)}} \sum_{b=1}^{d_{B(N_A)}} \vec{z}_b(N_A) \vec{z}_b(N_A)^T.
\end{equation}
Since $\vec{z}_b(N_A)$ is a different random vector for every $b$, the ranks of the outer-product matrices in Eq.~(\ref{def_I_M}) add up and result in ${\rm Min}\left[ d_{B(N_A)}, d_{N_A} \right]$ eigenvalues that are nonzero. Hence, there will be $d_{N_A} - d_{B(N_A)}$ zero eigenvalues in blocks with $N_A > N_A^*$ particles, where $N_A^*$ is the lowest $N_A$ for which $d_{B(N_A)} < d_{N_A}$. This analysis reveals that, in the regime $N<L_A$, a large fraction of the eigenvalues of $I + M(N_A)$ (for $N_A > N_A^*$) can be zero. They do not contribute to the entanglement entropy. As a result, the average of the nonzero eigenvalues $\bar\Lambda_{N_A} = d_{N_A}/d_{B(N_A)}$ can be very large and can, as we show later, give rise to a contribution that grows with the volume of the subsystem.

While the actual values of the nonzero eigenvalues $\Lambda_j$ are in general unknown, our analysis so far implies that, when $N_A > N_A^*$, the average fluctuation term of the entanglement entropy equals $\bar S_{\rm fluct}(N_A) = - \bar \lambda_{N_A} \sum_{j=1}^{d_{B(N_A)}} \overline{\Lambda_{j} \ln \Lambda_{j}}$. One can set an upper bound to the fluctuation term, $\bar S_{\rm fluct} \leq S^{\rm bound}_{\rm fluct}$, by replacing $\Lambda_j$ with $\bar\Lambda_{N_A}$, which yields
\begin{equation} \label{def_Sbound}
 S^{\rm bound}_{\rm fluct} = - \sum_{N_A=N_A^*}^N d_{N_A} \bar \lambda_{N_A} \ln \bar\Lambda_{N_A} \, .
\end{equation}
Note that, despite the similarity between Eq.~\eqref{def_Sbound} and Eq.~(\ref{Smf_long}), $S^{\rm bound}_{\rm fluct}\leq0$ ($\bar\Lambda_{N_A} > 1$) while $S_{\rm MF}\geq0$. Note that $S^{\rm bound}_{\rm fluct} = 0$ if $L_A < N$.

Summarizing our results so far, we have derived an analytic expression for the upper bound of the average entanglement entropy of random canonical states with a fixed particle number
\begin{equation} \label{def_S_upper}
 \bar S \leq S_{\rm MF} + S^{\rm bound}_{\rm fluct} \, ,
\end{equation}
where $S_{\rm MF}$ and $S^{\rm bound}_{\rm fluct}$ are given by Eqs.~(\ref{Smf_long}) and~(\ref{def_Sbound}), respectively \footnote{Notice that $S_{\rm MF}$ valid for any $N$, while $S^{\rm bound}_{\rm fluct}$ is only valid for $N\leq L/2$.}. The key final step is to find how $S^{\rm bound}_{\rm fluct}$ scales with the subsystem size $L_A$. Here, we focus on $f=1/2$ and derive a closed-form expression of the two leading terms in the limit of large system sizes~\cite{suppmat}
\begin{align} \label{def_Sstar}
 S^{\rm bound*}_{\rm fluct}  = & - \sqrt{L_A} \, \ln\left(\frac{1-n}{n}\right) \sqrt{ \frac{n(1-n)}{\pi}} \nonumber \\ & + \frac{1}{\sqrt{L_A}} \, \frac{(1-2n)}{3\sqrt{\pi n (1-n)}} \, .
\end{align}
Hence, for $n<1/2$, $S_{\rm MF} + S^{\rm bound}_{\rm fluct}$ is fundamentally different from Eq.~(\ref{def_Page}), which predicts an $O(1)$ correction to the mean-field entanglement entropy $S_{\rm MF}$. (Note that, for $n=1/2$, $S^{\rm bound*}_{\rm fluct}=0$.) Moreover, $S_{\rm MF} + S^{\rm bound}_{\rm fluct}$ is also fundamentally different from the entropy in the canonical ensemble at infinite temperature, $S_A = \ln {\cal D}_N/2 \approx -L_A \left[ n \ln n + (1-n) \ln(1-n) \right] - (1/4)\ln (L_A) + C_n$, for which the largest correction to the mean-field entanglement entropy is logarithmic in $L_A$.

In Ref.~\cite{suppmat}, we provide numerical evidence that $S^{\rm bound}_{\rm fluct}$ still grows as $\sqrt{L_A}$ for subsystem sizes that satisfy $\lim_{L\to\infty}(L/2-L_A)/\sqrt{L} = {\rm const}$. On the other hand, if $L_A = f L$ with $f < 1/2$, $S^{\rm bound}_{\rm fluct}$ decreases exponentially fast with increasing $L_A$.

Figure~\ref{fig2} shows the rescaled entanglement entropy $(\bar S - S_{\rm MF})/\sqrt{L_A}$ when $f=1/2$, for different average site occupations $n=1/3$, $1/4$ and $1/6$. We compare exact numerical results for random canonical states, Eq.~(\ref{def_psiN}), and for eigenstates of the Hamiltonian in Eq.~(\ref{def_Ham}), with the predictions from $S^{\rm bound}_{\rm fluct}$ and $S^{\rm bound*}_{\rm fluct}$ in Eqs.~(\ref{def_Sbound}) and~(\ref{def_Sstar}), respectively. In all cases, as expected, $S^{\rm bound}_{\rm fluct}/\sqrt{L_A}$ provides an upper bound. Remarkably, linear extrapolations of the numerical results (shown only for the random canonical states) appear to saturate the bound for $L_A \to \infty$. This indicates that, in Eq.~(\ref{def_S_upper}), the equality in likely to hold in the thermodynamic limit.

{\it Random states without a fixed particle number.--} To conclude, we provide numerical evidence that the main results derived for states with a fixed particle number remain valid for states without a fixed particle number. We consider random states
\begin{equation} \label{def_psi_Nall}
 |\psi'\rangle = \sum_{m=1}^{\cal D} \frac{z_m}{\sqrt{Z}} \, e^{\mu \hat N/2} |m\rangle \, ,
\end{equation}
where $z_m$ is a normally distributed real random number with zero mean and variance one, $|m\rangle$ is a base ket in the site-occupation basis, ${\cal D} = 2^L$, and $Z = (1+e^\mu)^L$. The chemical potential $\mu=\ln [n/(1-n)]$ sets the average site occupation to $n$.

\begin{figure}[!t]
\begin{center}
\includegraphics[width=0.99\columnwidth]{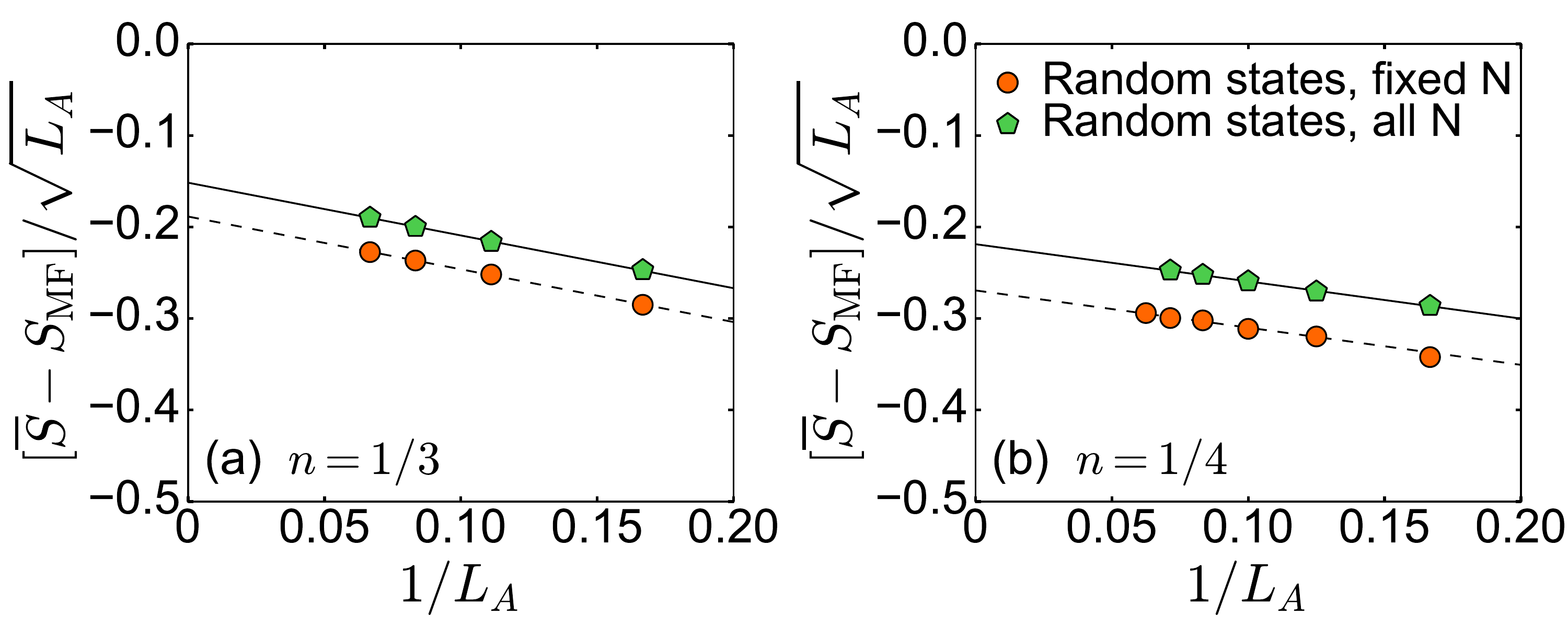}
\vspace*{-0.45cm}
\caption{Entanglement entropy of random pure states when $f=1/2$, for average site occupations $n=1/3$ (a) and $n=1/4$ in (b). The circles display results for random canonical states, Eq.~(\ref{def_psiN}). [Circles and dashed lines are identical to the ones in Figs.~\ref{fig2}(a) and~\ref{fig2}(b).] The pentagons display results for random states without a fixed particle number, Eq.~(\ref{def_psi_Nall}), where $S_{\rm MF} \to S_{\rm MF}'$. Solid lines are linear fits in which the slope is chosen to be identical to the one for the dashed lines (see Ref.~\cite{suppmat} for details on the numerical calculations).}
\label{fig3}
\end{center}
\end{figure}

Following the same procedure as for states with a fixed particle number [see Eq.~(\ref{def_psiN})], we obtain the analog of $S_{\rm MF}$ in Eq.~(\ref{def_Smf}) for the states in Eq.~\eqref{def_psi_Nall} (see Ref.~\cite{suppmat})
\begin{equation} \label{def_Smf_Nall}
 S_{\rm MF}' = -L_A \left[ n \ln n + (1-n) \ln(1-n) \right] \, ,
\end{equation}
which is, for nonzero $f$, larger than $S_{\rm MF}$ in Eq.~(\ref{Smf_short}). In Fig.~\ref{fig3}, we compare the rescaled average entanglement entropy for both classes of random states. One can see that, in both cases, the fluctuation contribution to the entanglement entropy $\bar S_{\rm fluct}$ is consistent with a $\sqrt{L_A}$ dependence as $L_A\rightarrow\infty$. The values of $\bar S_{\rm fluct}$ are larger for states without a fixed particle number, consistent with the results obtained for the mean-field entropies.

{\it Discussion.--}
Using random pure states with a fixed particle number and normally distributed real coefficients, we studied the deviation of the average entanglement entropy from the maximal value introduced by the fluctuations of the matrix elements of the reduced density matrix. For $f=1/2$ and $n\neq1/2$, we proved that there is a lower bound to that deviation that grows as the square root of the subsystem volume. Exact numerical results for random canonical states and for highly excited eigenstates of a quantum chaotic model indicate that the bound is saturated as $L_A\rightarrow\infty$. We also presented numerical evidence suggesting that qualitatively similar results hold in the absence of the fixed particle number constraint, for which analytic results are not available. Our results show that while the leading term for the average entanglement entropy of high-energy eigenstates of quantum-chaotic Hamiltonians is likely the maximal entanglement entropy, the correction for $f=1/2$ and $n\neq1/2$ must at least grow with the square root of the volume of the subsystem (as opposed to be a constant, as dictated by Page's result).
Our results highlight that, away from half filling, high-energy eigenstates of quantum chaotic Hamiltonians are not like typical pure states in the Hilbert space. They also make apparent that large deviations from the maximal entanglement entropy can be observed in experiments and numerical calculations in finite systems.

{\it Acknowledgments.}
This work was supported by the Office of Naval Research, Grant No.~N00014-14-1-0540. We acknowledge discussions with Lucas Hackl and Eugenio Bianchi. The computations were done at the Institute for CyberScience at Penn State.


\bibliographystyle{biblev1}
\bibliography{references}

\newpage
\phantom{a}
\newpage
\setcounter{figure}{0}
\setcounter{equation}{0}

\renewcommand{\thetable}{S\arabic{table}}
\renewcommand{\thefigure}{S\arabic{figure}}
\renewcommand{\theequation}{S\arabic{equation}}

\renewcommand{\thesection}{S\arabic{section}}

\onecolumngrid

\begin{center}

{\large \bf Supplemental Material:\\
Entanglement Entropy of Eigenstates of Quantum Chaotic Hamiltonians}\\

\vspace{0.3cm}

Lev Vidmar and Marcos Rigol\\
{\it Department of Physics, The Pennsylvania State University, University Park, PA 16802, USA}

\end{center}

\vspace{0.6cm}

\twocolumngrid

\label{pagesupp}

\section{Normalization and numerical implementation of random states} \label{secS1}

We note that the random states in Eqs.~(\ref{def_psiN}) and~(\ref{def_psi_Nall}) in the main text are not exactly normalized in finite systems. The normalized state corresponding to Eq.~(\ref{def_psiN}) can be written as
\begin{equation}
 |\tilde{\psi}_N \rangle = \frac{1}{\sqrt{\cal N}} \sum_{j=1}^{{\cal D}_N} \frac{z_j}{\sqrt{{\cal D}_N}} |j\rangle \, ,
\end{equation}
where the normalization factor ${\cal N}$ equals
\begin{equation}
 {\cal N} = \frac{1}{{\cal D}_N} \sum_{j=1}^{{\cal D}_N} z_j^2 \, .
\end{equation}
Here, $z_j^2$ are random numbers with mean $\eta = 1$ and variance $\sigma_\eta^2 = 2$. For large Hilbert-space dimensions ${\cal D}_N$, one can apply the central limit theorem, which yields the average $\overline{\cal N} = \eta = 1$ and the variance $\sigma_{\cal N}^2 = \sigma_\eta^2/{{\cal D}_N}$. Hence, the fluctuations of ${\cal N}$ decay as $1/{\cal D}_N$, i.e., exponentially fast with increasing the number of lattice sites $L$ for average particle occupations $0<n<1$ [see Eq.~(\ref{def_dn})]. We therefore use Eq.~(\ref{def_psiN}) in all analytical calculations. Similar considerations apply to Eq.~(\ref{def_psi_Nall}).

In the numerical calculations in finite systems, we normalize each random state so that the traces of the density matrices are exactly one. For Figs.~\ref{fig1}-\ref{fig3}, we calculate the entanglement entropy $\bar S$ averaged over $N_r$ random states. Filled symbols in the main panel of Fig.~\ref{fig1} (random canonical states) show $\bar S$ averaged over $N_r=1000$ realizations. In the inset (results for $f=n=1/2$), we average over $N_r=1000$ realizations for $L \leq 26$, and over $N_r=100$ realizations for $L > 26$. Circles in Figs.~\ref{fig2} and~\ref{fig3} (random canonical states) show $\bar S$ averaged over $N_r=1000$ realizations for $L \leq 30$, and over $N_r=100$ realizations for $L> 30$. Pentagons in Fig.~\ref{fig3} (random states without a fixed particle number) show $\bar S$ averaged over $N_r = 1000$ (or $N_r=10$) realizations for $L \leq 24$ (or $L > 24$). The standard deviation of all averages is smaller than the symbol sizes in all figures.

\section{Derivation of Eq.~(\ref{Smf_short})} \label{sec_Smf_short}

Here, we derive $S_{\rm MF}^*$ in Eq.~(\ref{Smf_short}), which is a closed-form expression for the mean-field entanglement entropy
\begin{equation} \label{Smf_long2}
S_{\rm MF} = - \sum_{N_A=N_A^{\rm min}}^{N_A^{\rm max}} d_{N_A} \bar\lambda_{N_A} \ln \bar\lambda_{N_A}
\end{equation}
introduced in Eqs.~(\ref{def_Smf}) and~(\ref{Smf_long}) in the main text. There, we also defined $N_A^{\rm min} = {\rm Max}[0,N-(L-L_A)]$, $N_A^{\rm max} = {\rm Min}[N,L_A]$, $d_{N_A} = \binom{L_A}{N_A}$, $\bar\lambda_{N_A} = d_{B(N_A)}/{\cal D}_N$, $d_{B(N_A)} = \binom{L-L_A}{N-N_A}$, and ${\cal D}_N = \binom{L}{N}$. We apply the Stirling's approximation $n! = \sqrt{2\pi n} (n/e)^n$ to simplify the binomial coefficients. For example, the Stirling's approximation for the Hilbert-space dimension ${\cal D}_N$ yields
\begin{equation} \label{def_dn}
 {\cal D}_N = \binom{L}{N} = \frac{1}{\sqrt{2\pi L} \sqrt{n(1-n)}} \left[ n^n \, (1-n)^{1-n} \right]^{-L} \, .
\end{equation}
Analogous expressions can be obtained for $d_{N_A}$ and $d_{B(N_A)}$. The derivation consists of two main steps: (i) rewrite $\ln \bar\lambda_{N_A}$, and (ii) express $d_{N_A} \bar\lambda_{N_A}$ as a Gaussian function.

In the first step, we rewrite $-\ln\bar\lambda_{N_A}$ as
\begin{align} \label{eq_logNa}
 - \ln & \bar\lambda_{n_A} = S_{\rm MF}' + \frac{\ln(1-f)}{2}  \nonumber \\ & + L_A(n_A-n) \ln\left(\frac{1-n}{n}\right) +  \frac{1}{2} \left( \ln \alpha_1 + \ln \alpha_2 \right) \nonumber \\ & + L(1-f) \left[ n \alpha_1 \ln \alpha_1 + (1-n) \alpha_2 \ln\alpha_2 \right] \, ,
\end{align}
where $S_{\rm MF}' = -L_A [n\ln n + (1-n)\ln(1-n)]$ is Eq.~(\ref{def_Smf_Nall}) in the main text, $\alpha_1 = 1 - \frac{(n_A-n)}{n} \frac{f}{1-f}$, $\alpha_2 = 1 + \frac{(n_A-n)}{1-n} \frac{f}{1-f}$, and $n_A = N_A/L_A$ ($\alpha_1, \alpha_2 \in [0,2]$). Equation~(\ref{eq_logNa}) can be seen to contain an $N_A$-independent part (the first two terms in the right hand side) and an $N_A$-dependent part (the last three terms in the right hand side). The $N_A$-independent part remains unchanged in the final expression of the entanglement entropy $S_{\rm MF}^*$, because of the trace normalization $\sum_{N_A} d_{N_A} \bar\lambda_{N_A} = 1$. All the $N_A$-dependent terms in Eq.~(\ref{eq_logNa}) can be expressed as power series of $(n_A-n)$, which we exploit in what follows.

In the second step, we rewrite $d_{N_A} \bar\lambda_{N_A}$. We first evaluate $\ln(d_{N_A} \bar\lambda_{N_A})$ in the same manner as $\ln\bar\lambda_{N_A}$ in Eq.~(\ref{eq_logNa}), and then expand the resulting expression in power series of $(n_A-n)$. This results in
\begin{align}
 \ln(d_{N_A} \bar\lambda_{N_A}) = & - \frac{L_A}{2(1-f) n (1-n)} (n_A - n)^2  \\ &- \frac{1}{2} \ln \left[ 2\pi L_A (1-f)n (1-n) \right] + \ldots \, .\nonumber
\end{align}
By introducing a new variable $\bar n = L_A (n_A-n)$, we express $d_{\bar n} \bar\lambda_{\bar n}$ as a Gaussian function
\begin{equation} \label{def_dlambda_gauss}
 d_{\bar n} \bar\lambda_{\bar n} = \frac{1}{\sqrt{2\pi}\sigma} e^{-\frac{\bar n^2}{2 \sigma^2}} \,
\end{equation}
with $\sigma = \sqrt{L_A (1-f)n(1-n)}$. This function is peaked at $n_A = n$, which sets the particle sector that has the maximal contribution to $S_\text{MF}$.

For $L_A \gg 1$, one can replace the sum in Eq.~(\ref{Smf_long2}) by an integral, $\sum_{N_A} \to \int_{-\infty}^\infty {\rm d}\bar n$, and hence $S_{\rm MF} = - \int_{-\infty}^\infty {\rm d}\bar n \, d_{\bar n} \bar\lambda_{\bar n} \ln \bar\lambda_{n_A}$. Since $ \ln \bar\lambda_{n_A}$ is a power series of $(n_A-n)$, with nonzero contributions only from terms with even powers, one has to evaluate a general integral of the form
\begin{align}
 I_m & =  \int_{-\infty}^\infty {\rm d}\bar n \, d_{\bar n} \bar\lambda_{\bar n} (n_A - n)^{2m} \nonumber \\ & = (2m-1)!! \left[ \frac{(1-f)n(1-n)}{L_A} \right]^m \,\, .
\end{align}
Hence, most of the terms in the second and third line in Eq.~(\ref{eq_logNa}) yield a contribution $O(1/L_A)$ or smaller to $S_{\rm MF}$. The only $O(1)$ contributions come from the first terms in the expansion of $\alpha_1 \ln\alpha_1$ and $\alpha_2 \ln\alpha_2$ in Eq.~(\ref{eq_logNa}), and yield $f/2$. The closed-form expression for the mean-field entanglement entropy then reads
\begin{equation} \label{Smf_short2}
 S_{\rm MF}^* = S_{\rm MF}' + \frac{f+\ln(1-f)}{2} + O(L_A^{-1}) \, ,
\end{equation}
which is Eq.~(\ref{Smf_short}) in the main text.

\section{Derivation of Eq.~(\ref{def_Sstar})} \label{sec_Sstar}

Here, we derive Eq.~(\ref{def_Sstar}) in the main text from
\begin{equation} \label{def_Sbound2}
 S_{\rm fluct}^{\rm bound} = -\sum_{N_A = N_A*}^N d_{N_A} \bar\lambda_{N_A} \ln \bar\Lambda_{N_A} \, ,
\end{equation}
where $\bar\Lambda_{N_A} = d_{N_A}/d_{B(N_A)}$. We set $f=L_A/L=1/2$, which implies that $N_A^*$ is the integer part of $N/2+1$. ($N_A^*$ is the lowest $N_A$ for which $d_{B(N_A)} < d_{N_A}$.)

The derivation is analogous to the one presented in Sec.~\ref{sec_Smf_short}. The term $d_{N_A} \bar\lambda_{N_A}$ is approximated by the Gaussian function $d_{\bar n} \bar\lambda_{\bar n}$ in Eq.~(\ref{def_dlambda_gauss}). The term $\ln \bar\Lambda_{N_A}$ is rewritten as
\begin{align} \label{eq_LlogNa}
 \ln & \bar\Lambda_{n_A}  = L (n_A-n) \ln \left( \frac{1-n}{n} \right) \nonumber \\ & - \frac{L}{2} \left[ n \ln \left( \frac{\beta_1^+}{\beta_1^-} \right) + (n_A-n) \ln(\beta_1^+ \beta_1^-) \right] \nonumber \\ & + \frac{L}{2} \left[ (1-n) \ln \left( \frac{\beta_2^+}{\beta_2^-} \right) + (n_A-n) \ln(\beta_2^+ \beta_2^-) \right] \, ,
\end{align}
where $\beta_1^{\pm} = 1 \pm (n_A-n)/n$ and $\beta_2^{\pm} = 1 \pm (n_A-n)/(1-n)$. Equation~(\ref{eq_LlogNa}) can be expanded in power series of $(n_A-n)$ [with nonzero coefficients only for odd powers of $(n_A-n)$]. We exploit this to obtain the final result.

For $L_A \gg 1$, and introducing $\bar n = L_A (n_A-n)$ as in Sec.~\ref{sec_Smf_short}, the sum in Eq.~(\ref{def_Sbound2}) can be replaced by the integral $S_{\rm fluct}^{\rm bound} = -\int_{0}^{\infty} {\rm d}\bar n \, d_{\bar n} \bar\lambda_{\bar n} \ln \bar\Lambda_{n_A}$. The general integral one needs to calculate in this case is
\begin{align}
 J_p = \int_0^\infty {\rm d}\bar n \, d_{\bar n} \bar\lambda_{\bar n} (n_A-n)^{2p+1} = \frac{p! \, 2^p}{\sqrt{2\pi}} \left[ \frac{n(1-n)}{2L_A} \right]^{p+1/2}.
\end{align}
The leading contribution to $S_{\rm fluct}^{\rm bound}$ comes from the first term on the right-hand side of Eq.~(\ref{eq_LlogNa}), and is proportional to $\sqrt{L_A}$. All other terms in Eq.~(\ref{eq_LlogNa}) yield a contribution that decays as $1/\sqrt{L_A}$ or faster. The closed-form expression for $S^{\rm bound}_{\rm fluct}$ that contains the two dominant terms is
\begin{align} \label{def_Sstar2}
 S^{\rm bound*}_{\rm fluct}  = & - \sqrt{L_A} \, \ln\left(\frac{1-n}{n}\right) \sqrt{ \frac{n(1-n)}{\pi}} \nonumber \\ & + \frac{1}{\sqrt{L_A}} \, \frac{(1-2n)}{3\sqrt{\pi n (1-n)}} + O(L_A^{-3/2}) \, ,
\end{align}
which is Eq.~(\ref{def_Sstar}) in the main text.

\begin{figure}[!t]
\begin{center}
\includegraphics[width=0.99\columnwidth]{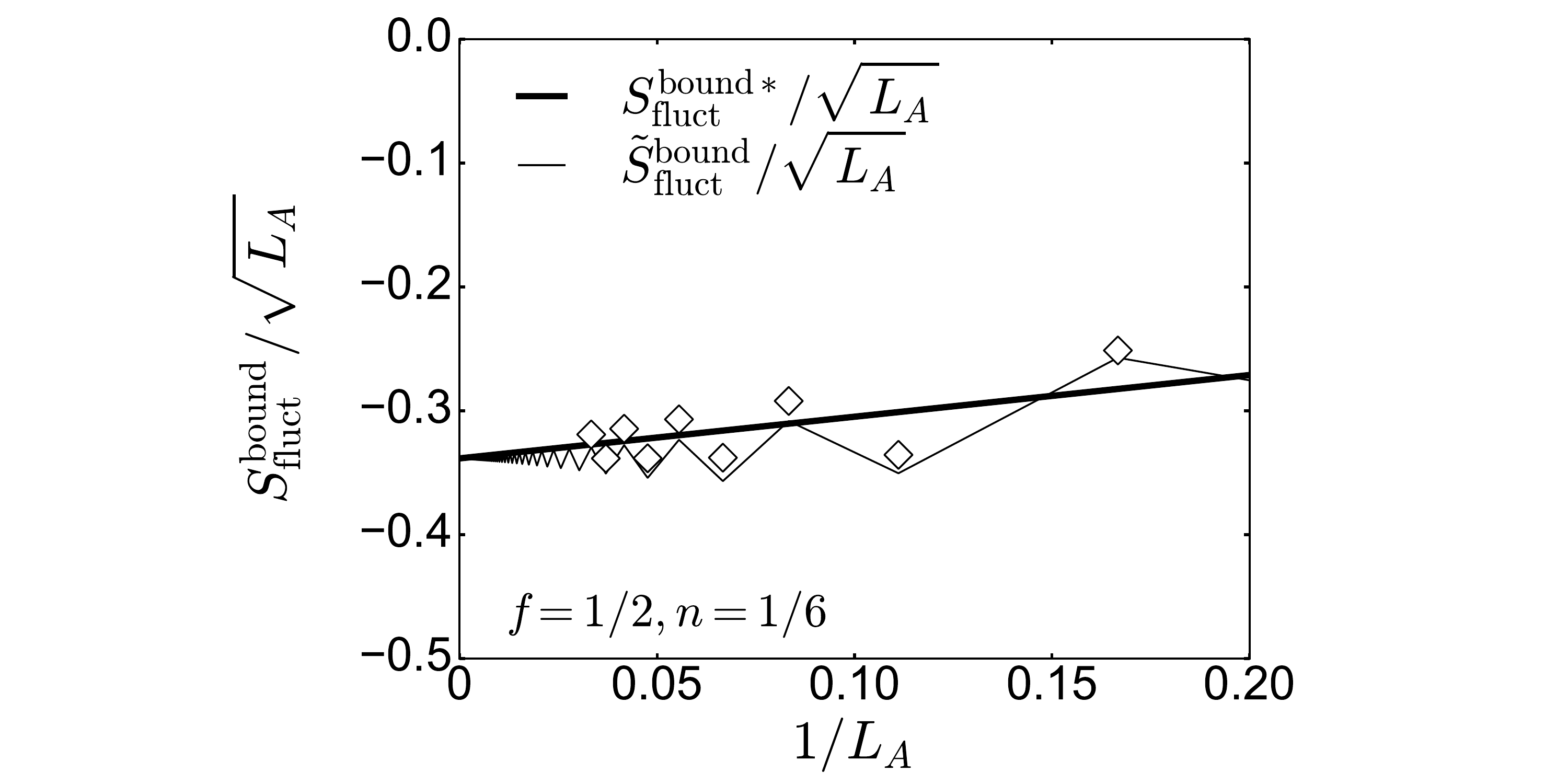}
\vspace*{-0.6cm}
\caption{Scaling of $S_{\rm fluct}^{\rm bound} / \sqrt{L_A}$ with the inverse subsystem size $1/L_A$ for $f=1/2$ and $n=1/6$. Diamonds are exact numerical results, the thick solid line is $S_{\rm fluct}^{\rm bound*}/\sqrt{L_A}$ from Eq.~(\ref{def_Sstar2}), and the thin solid line is $\tilde{S}_{\rm fluct}^{\rm bound}/\sqrt{L_A}$. [We obtain $\tilde{S}_{\rm fluct}^{\rm bound}$ from $S_{\rm fluct}^{\rm bound}$ upon replacing the binomial coefficients in Eq.~(\ref{def_Sbound2}) with Stirling's approximation.]
}
\label{figS1}
\end{center}
\end{figure}

Figure~\ref{figS1} shows the results at $f=1/2$ and $n=1/6$. Diamonds display the exact numerical values of $S_{\rm fluct}^{\rm bound}$ [Eq.~(\ref{def_Sbound2})] and the thick solid line is $S_{\rm fluct}^{\rm bound*}$ in Eq.~(\ref{def_Sstar2}) [same results as in Fig.~\ref{fig2}(c) in the main text]. Moreover, we also plot $\tilde{S}_{\rm fluct}^{\rm bound}$, which is obtained numerically from $S_{\rm fluct}^{\rm bound}$ upon replacing the binomial coefficients in Eq.~(\ref{def_Sbound2}) with their Stirling approximation (thin solid line). The results show a perfect agreement as $1/L_A \to 0$. In finite systems, the exact results for $S_{\rm fluct}^{\rm bound}$ exhibit an even-odd effect, which can also be seen for $n=1/4$, but is absent for $n=1/3$ (see Fig.~\ref{fig2} in the main text). The origin of this behavior is that $N/2$ may be noninteger for $n=1/4$ and $1/6$, and, hence, $N_A^*$ is identical for two consecutive subsystem sizes $L_A$ that are compatible with a given $n$. This is not the case for $n=1/3$ for which $N$ is always even.

\subsection{Subsystem fractions $f\neq1/2$}

The derivation of the closed-form expression for $S^{\rm bound}_{\rm fluct}$ presented above is for the case $L_A = f L$, with $f=1/2$, where a simple expression for $N_A^*$ is available for all $L$. For other subsystem sizes, we calculate $N_A^*$ numerically and study the scaling of $\tilde{S}^{\rm bound}_{\rm fluct}$ with $L$. (Note that for nonzero $n$ and $f$, $\tilde{S}^{\rm bound}_{\rm fluct} \to S^{\rm bound}_{\rm fluct}$ in the limit $L\to\infty$.)

\begin{figure}[!h]
\begin{center}
\includegraphics[width=0.99\columnwidth]{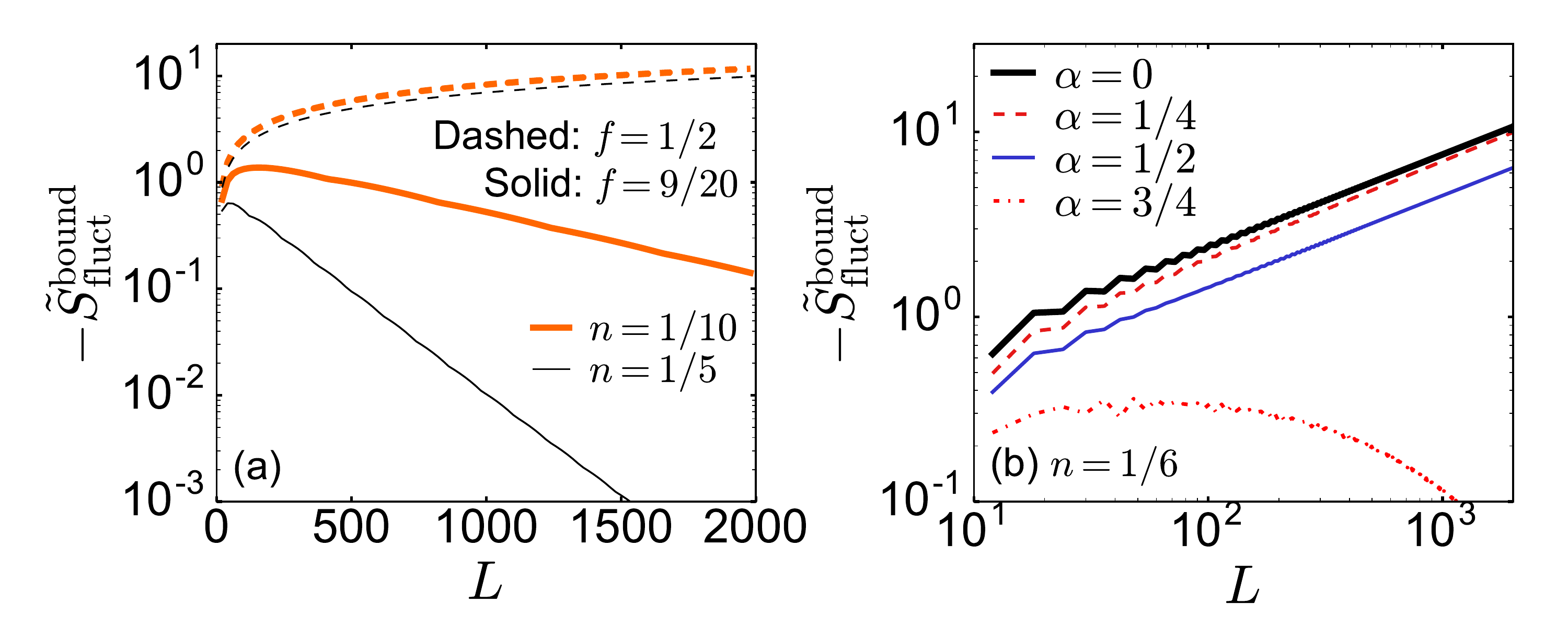}
\vspace*{-0.6cm}
\caption{Scaling of $-\tilde{S}_{\rm fluct}^{\rm bound}$ vs $L$.
(a)
Subsystem volumes $L_A = f L$ for $f=1/2$ (dashed lines) and $f=9/20$ (solid lines). Results are shown for $n=1/10$ (thick red lines) and for $n=1/5$ (thin black lines).
(b) 
Subsystem volumes $L_A = L/2 - \gamma L^\alpha$ and $n=1/6$, for different values of $\alpha$.
For $\alpha = 0$, we set $\gamma = 0$, while for $\alpha>0$ we set $\gamma=1/4$.
}
\label{figS2}
\end{center}
\end{figure}

Figure~\ref{figS2}(a) shows results for a subsystem volume $L_A = f L$ where $f$ is close to, but smaller than, one half. We chose $f=9/20$ and focus on small $n$ (the absolute value of $S_{\rm fluct}^{\rm bound}$ increases with decreasing $n$, for $n$ not too small). In both cases considered in Fig.~\ref{figS2}(a), $n=1/10$ and $n=1/5$, we observe an exponential decay with increasing $L$ (for large values of $L$). These results suggest that $S_{\rm fluct}^{\rm bound}$ vanishes exponentially fast with increasing $L$ for bipartitions with $f \neq 1/2$. In Fig.~\ref{figS2}(a), we also show results for $f=1/2$, which are in stark contrast with those for $f=9/20$ as, for $f=1/2$, $S_{\rm fluct}^{\rm bound}$ grows with increasing $L$.

Next, we study subsystem sizes $L_A$ that are not proportional to $L$. We focus on $n=1/6$ and $L_A = L/2 - \gamma L^\alpha$, where $\alpha < 1$. In these cases, the subsystem fraction $f=1/2$ is restored only in the limit $L\to \infty$. The results in Fig.~\ref{figS2}(b) show that: (i) if $\alpha < 1/2$, $S_{\rm fluct}^{\rm bound}$ approaches the same value as for $L_A = L/2$, (ii) if $\alpha = 1/2$, $S_{\rm fluct}^{\rm bound}$ still increases subextensively, however, with a smaller prefactor, and (iii) if $\alpha > 1/2$, $S_{\rm fluct}^{\rm bound}$ vanishes with increasing system size.

Our results therefore suggest that the subextensive scaling of $S_{\rm fluct}^{\rm bound}$ with $L$ can be observed for subsystem volumes $L_A$ that satisfy $\lim_{L\to\infty} (L/2-L_A)/\sqrt{L} = \mbox{const.}$ The function $S_{\rm fluct}^{\rm bound}(n,L_A)$ can then be written as
\begin{equation}
 S_{\rm fluct}^{\rm bound}(n, L_A, L) \simeq G\left(n,\frac{L/2 - L_A}{\sqrt{L}} \right) \; S_{\rm fluct}^{\rm bound*} \, ,
\end{equation}
where $S_{\rm fluct}^{\rm bound*}$ is evaluated at $L/2$. By defining $x = (L/2 - L_A)/\sqrt{L}$, the function $G(n,x)$ scales in the limit $L \to \infty$ as: (i) $G(n,x) \to 1$ if $x \to 0$, (ii) $G(n,x) \to c_1(n)$, with $0 < c_1 < 1$, if $x$ approaches a nonzero constant, and (iii) $G(n,x) \to 0$ if $x \to \infty$.

\section{Derivation of Equation~(\ref{def_Smf_Nall})}

Here we derive the reduced density matrix for the random states $|\psi'\rangle$ in Eq.~(\ref{def_psi_Nall}) without a fixed particle number. We use the same notation as for the random canonical states in the main text. The reduced density matrix of the subsystem $A$, defined as $\hat\rho_A = {\rm Tr}_B\{ |\psi'\rangle \langle \psi' | \}$, is
\begin{align}
 \hat \rho_A = \sum_{N_A,N_{A}' = 0}^{L_A} & \sum_{a= 1}^{d_{N_A}} \sum_{a'= 1}^{d_{N_A'}} |a, N_A\rangle \langle a', N_A'| \nonumber \\ & \times \frac{e^{\mu (N_A + N_{A}')/2}}{Z} \, C(a,a',N_A,N_A') \, ,
\end{align}
where
\begin{align}
 C(a,a',&N_A,N_A') = \nonumber \\ & \sum_{N_B = 0}^{L-L_A} \sum_{b=1}^{d_B} z_{a,b}(N_A,N_B) z_{a',b} (N_A',N_B) \, e^{\mu N_B} \, ,
\end{align}
$\mu = \ln[n/(1-n)]$ and $d_B = \binom{L-L_A}{N_B}$. The average of the reduced density matrix over different random states is obtained using the relation $\overline{C(a,a',N_A,N_A')} = (1+e^\mu)^{L-L_A} \, \delta_{a,a'} \delta_{N_A,N_A'}$. It reads
\begin{equation}
 \hat{\overline{\rho}}_A = \sum_{N_A=0}^{L_A} \sum_{a=1}^{d_{N_A}} | a, N_A\rangle \langle a, N_A | \frac{e^{\mu N_A}}{(1+e^{\mu})^{L_A}} \, .
\end{equation}
Using the latter expression, and the relation between $\mu$ and $n$, it follows straightforwardly that the mean-field contribution to the entanglement entropy $S_{\rm MF}'$, as shown in Eq.~(\ref{def_Smf_Nall}) in the main text, is
\begin{equation}
 S_{\rm MF}' = -L_A \left[ n \ln n + (1-n) \ln(1-n) \right] \, .
\end{equation}

\end{document}